\documentclass[prd,a4,twocolumn,superscriptaddress,nofootinbib]{revtex4}
\usepackage{amsmath,amssymb,epsfig,natbib}
\usepackage{hyperref}
\usepackage{ifthen}

\begin{document}

\renewcommand{\bar}[1]{#1}
\newcommand{\nHI}{{n_{HI}}}
\newcommand{\clh}{\mathcal{H}}
\newcommand{\ud}{{\rm d}}
\renewcommand{\eprint}[1]{\href{http://arxiv.org/abs/#1}{#1}}
\newcommand{\adsurl}[1]{\href{#1}{ADS}}
\renewcommand{\bibinfo}[2]{\ifthenelse{\equal{#1}{isbn}}{%
\href{http://cosmologist.info/ISBN/#2}{#2}}{#2}}
\newcommand{\vort}{\varpi}
\newcommand\ba{\begin{eqnarray}}
\newcommand\ea{\end{eqnarray}}
\newcommand\be{\begin{equation}}
\newcommand\ee{\end{equation}}
\newcommand\lagrange{{\cal L}}
\newcommand\cll{{\cal L}}
\newcommand\clx{{\cal X}}
\newcommand\clz{{\cal Z}}
\newcommand\clv{{\cal V}}
\newcommand\clo{{\cal O}}
\newcommand\cla{{\cal A}}
\newcommand{\uD}{{\mathrm{D}}}
\newcommand{\calE}{{\cal E}}
\newcommand{\calB}{{\cal B}}
\newcommand{\curl}{\,\mbox{curl}\,}
\newcommand\del{\nabla}
\newcommand\Tr{{\rm Tr}}
\newcommand\half{{\frac{1}{2}}}
\newcommand\fourth{{1\over 8}}
\newcommand\bibi{\bibitem}
\newcommand{\kf}{\beta}
\newcommand{\kfprod}{\alpha}
\newcommand\calS{{\cal S}}
\renewcommand\H{{\cal H}}
\newcommand\K{{\rm K}}
\newcommand\mK{{\rm mK}}

\newcommand\opacity{\tau_c^{-1}}

\newcommand{\Psil}{\Psi_l}
\newcommand{\bsigma}{{\bar{\sigma}}}
\newcommand{\bI}{\bar{I}}
\newcommand{\clk}{{\cal K}}

\newcommand{\bq}{\bar{q}}
\newcommand{\bv}{\bar{v}}
\renewcommand\P{{\cal P}}
\newcommand{\numfrac}[2]{{\textstyle \frac{#1}{#2}}}

\newcommand{\la}{\langle}
\newcommand{\ra}{\rangle}

\newcommand{\fHe}{f{_\text{He}}}
\newcommand{\Omtot}{\Omega_{\mathrm{tot}}}
\newcommand\xx{\mbox{\boldmath $x$}}
\newcommand{\phpr} {\phi'}
\newcommand{\gam}{\gamma_{ij}}
\newcommand{\sqgam}{\sqrt{\gamma}}
\newcommand{\delk}{\Delta+3{\K}}
\newcommand{\dph}{\delta\phi}
\newcommand{\om} {\Omega}
\newcommand{\dom}{\delta^{(3)}\left(\Omega\right)}
\newcommand{\rar}{\rightarrow}
\newcommand{\Rar}{\Rightarrow}
\newcommand\gsim{ \lower .75ex \hbox{$\sim$} \llap{\raise .27ex \hbox{$>$}} }
\newcommand\lsim{ \lower .75ex \hbox{$\sim$} \llap{\raise .27ex \hbox{$<$}} }
\newcommand\bigdot[1] {\stackrel{\mbox{{\huge .}}}{#1}}
\newcommand\bigddot[1] {\stackrel{\mbox{{\huge ..}}}{#1}}
\newcommand{\Mpc}{\text{Mpc}}
\newcommand{\Al}{{A_l}}
\newcommand{\Bl}{{B_l}}
\newcommand{\eAl}{e^\Al}
\newcommand{\ix}{{(i)}}
\newcommand{\ixp}{{(i+1)}}
\renewcommand{\k}{\beta}
\newcommand{\HD}{\mathrm{D}}

\newcommand{\nonflat}[1]{#1}
\newcommand{\Cgl}{C_{\text{gl}}}
\newcommand{\Cgltwo}{C_{\text{gl},2}}
\newcommand{\clp}{{\cal P}}
\newcommand{\He}{{\rm{He}}}
\newcommand{\Mhz}{{\rm MHz}}
\newcommand{\vx}{{\mathbf{x}}}
\newcommand{\ve}{{\mathbf{e}}}
\newcommand{\vetilde}{\tilde{\mathbf{e}}}
\newcommand{\vv}{{\mathbf{v}}}
\newcommand{\vk}{{\mathbf{k}}}

\newcommand{\vnhat}{{\hat{\mathbf{n}}}}
\newcommand{\vkhat}{{\hat{\mathbf{k}}}}
\newcommand{\taueps}{{\tau_\epsilon}}

\newcommand{\vgrad}{{\mathbf{\nabla}}}
\newcommand{\fbarln}{\bar{f}_{,\ln\epsilon}(\epsilon)}
\newcommand{\etaad}{\eta_a}
\title{Linear effects of perturbed recombination}

\author{Antony Lewis}
\homepage{http://cosmologist.info}

 \affiliation{Institute of Astronomy, Madingley Road, Cambridge, CB3 0HA, UK.}

\date{\today}

\begin{abstract}
Perturbations in the ionization fraction after recombination affect the Compton cooling of density perturbations. Once the gas temperature starts to decouple from the CMB temperature, ionization fraction perturbations can have a significant influence on the subsequent gas temperature perturbation evolution. This directly affects the 21cm spin temperature of the gas, and also modifies the small-scale baryon perturbation evolution via the difference in baryon pressure. The effect on the gas temperature perturbations can be significant on all scales, and galactic-scale baryon perturbations are modified at the percent level at redshifts $z\agt 100$ where numerical simulations are typically started.

\end{abstract}
\maketitle

\vskip .2in

\section{Introduction}

Perturbations in the ionization fraction have almost no effect on the CMB power spectrum. This is because an isotropic photon field scattering from an arbitrary static anisotropic electron distribution will still look isotropic. Perturbed recombination only changes the CMB though effects such as the scattering of the \emph{perturbed} temperature from the perturbed ionization fraction, which is a second-order process. However ionization fraction perturbations can influence the gas temperature perturbation evolution at linear order by changing the local Compton cooling rate. This in turn will modify the spin temperature perturbation relevant for dark-age 21cm observations~\cite{Scott90,Loeb:2003ya,Naoz:2005pd,Furlanetto:2006jb,Lewis:2007kz}, and will also change the Jeans'-scale baryon density evolution due to the difference in baryon pressure. This can be significant at redshifts $z \agt 100$ where numerical simulations are typically started, and should be included in any accurate calculation of the small-scale linear-theory baryon power spectrum at high redshift.

The post-recombination effect of ionization fraction perturbations can be understood as follows. After the main recombination epoch the temperature is well below the ionization energy of the lowest states of hydrogen. The photoionization rate is therefore very low, and electrons captured by protons result in the net production of neutral hydrogen atoms. Recombination is therefore limited by the low probability for electron capture in the sparse gas, and the recombination rate depends predominantly on the atomic density and velocity distribution (temperature). Over-dense regions, which are also hotter due to adiabatic contraction, will have more recombinations, and hence have a lower ionization fraction than the background. This is turn reduces the coupling to the CMB temperature via Compton scattering because there are fewer free electrons. After recombination the gas temperature is below the CMB temperature due to adiabatic cooling; the residual lower ionization fraction in an overdensity therefore reduces the coupling to the higher CMB temperature, and hence makes the gas temperature lower than it would have been. A lower gas temperature implies a lower baryon pressure, and hence growth of small-scale baryon perturbations is less suppressed by pressure support.

In this paper we give an approximate numerical analysis of perturbed recombination to quantify the linear effect on the gas temperature and small-scale baryon density evolution.  We calculate the evolution only where linear-theory is expected to be quite accurate (redshift $z \agt 30$). Note however that the overall accuracy of the calculation is currently limited by the understanding of the complex background recombination process (see e.g. Refs.~\cite{Chluba:2006bc,Chluba:2007yp,Switzer:2007sq}); a detailed accurate treatment of perturbed recombination at all redshifts will have to wait until the detailed physics of recombination is more fully understood.

\section{Perturbed recombination}

The main effect of perturbed recombination during the dark ages can be obtained straightforwardly by considering the ionization fraction evolution after the main recombination event. This is because the perturbations grow after recombination, and the result at redshifts of a few hundred is not very sensitive to the initial value at the end of recombination.
Defining an ionization fraction $x_e$, the number density of neutral hydrogen atoms is $n_{HI}=(1-x_e)n_H$, where $n_H$ is the total number density of ionized and unionized hydrogen. We work with conformal time (derivatives denoted by an over-dot) and consider only scalar perturbations in the synchronous gauge (so the CDM peculiar velocity is zero). We assume a standard flat concordance cold dark matter cosmology~\cite{Spergel:2006hy} in which the dark matter does not decay or annihilate (if it does the thermal history can be significantly altered~\cite{Furlanetto:2006wp}).

At late times the evolution of the background ionization fraction is given by
\begin{eqnarray}
\dot{x}_e &\approx & -a \alpha x_e^2 n_H,
\end{eqnarray}
where $a$ is the scale factor and $\alpha$ is the recombination coefficient, a function of the gas temperature $T_g$ fit by
\begin{equation}
\alpha = F   \frac{ a_\alpha (T_g/10^4 K)^b}{1+c  (T_g/10^4 K)^d} \text{m}^3\text{s}^{-1}.
\end{equation}
Here $F$ is a fudge factor taken to be $1.14$ from RECFAST~\cite{Seager:1999km} to calibrate the result to a multi-level atom calculation, and $a_\alpha = 4.309\times 10^{-19}$, $b=-0.6166$, $c=0.6703$, $d=0.5300$.

At late times the linear ionization fraction perturbation $\Delta_{x_e}$ then satisfies
\begin{eqnarray}
\dot{\Delta}_{x_e} &\approx& -a \alpha x_e n_H \left( \Delta_\alpha + \Delta_{x_e} +\Delta_{n_H}\right),
\label{dxe}
\end{eqnarray}
where $\Delta_i$ are the fractional perturbations and
\begin{eqnarray}
\Delta_\alpha &=& \frac{ b + c(T_g/10^4 K)^d(b-d)}{1+c(T_g/10^4 K)^d}\Delta_{T_g}.
\end{eqnarray}
Since the fractional baryon perturbation $\Delta_{n_H} \approx \Delta_{g}$ grows rapidly after recombination as the baryons fall in to the dark matter potential wells, the detailed evolution during recombination is not important for the late time amplitude of $\Delta_{x_e}$.   The sign of $\Delta_{x_e}$ is opposite to that of $\Delta_g$ because over-densities recombine more efficiently, so the ionization fraction is below that of the background at late times.

Note that the overall calculation is limited by the precision of the RECFAST model, in which a single fudge-factor accounts for deviations of an effective three level atom model from the full result. For example Ref.~\cite{Chluba:2006bc} find a $ \agt 5\%$ difference in  $x_e$ at $z\sim 200$ compared to the RECFAST model. Our results for the perturbation evolution are therefore of limited accuracy, though the relative importance of the ionization fraction perturbations should be roughly correct.

\section{Gas temperature evolution}

\begin{figure}
\begin{center}
\epsfig{figure=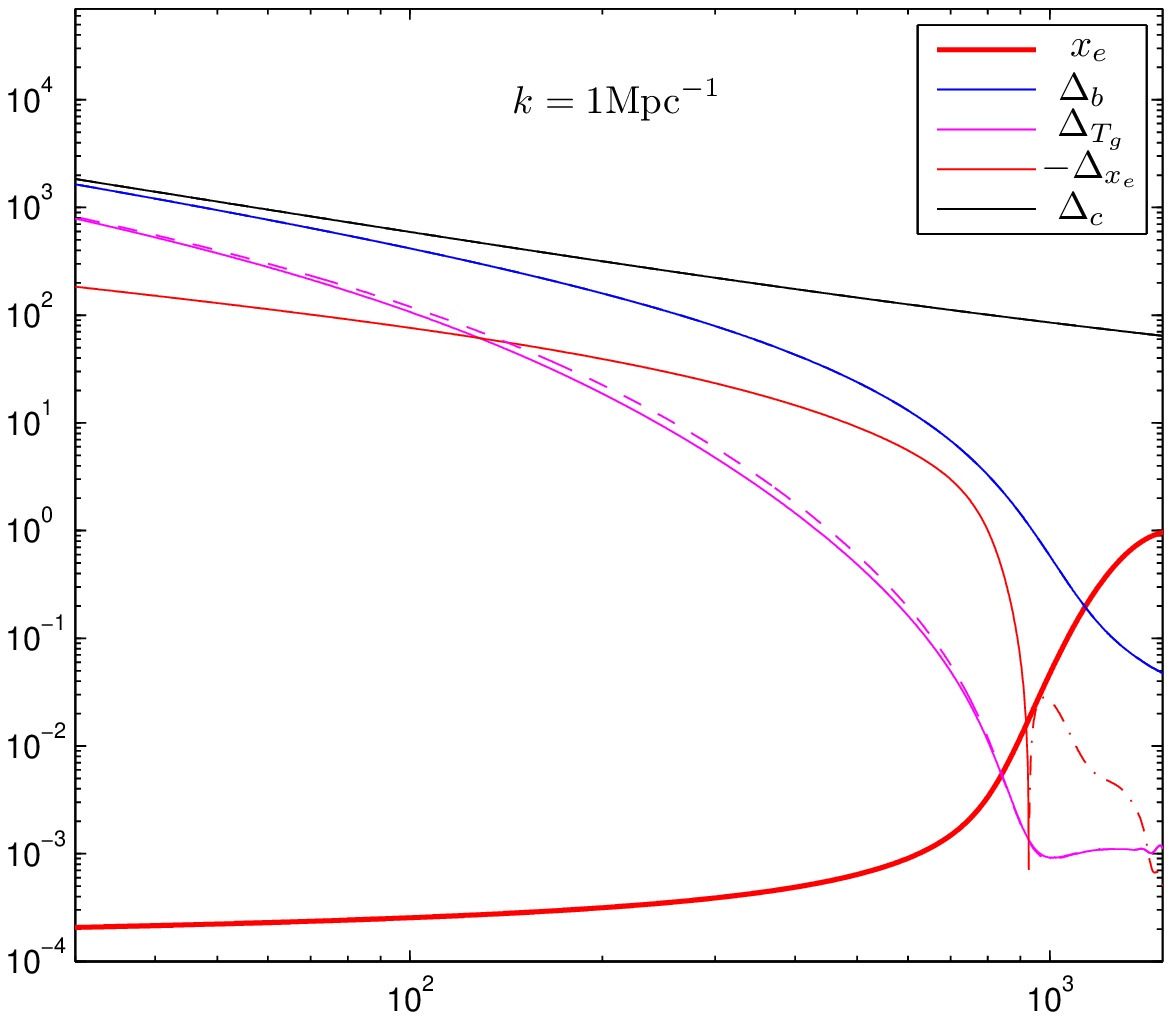,width=8cm}
\epsfig{figure=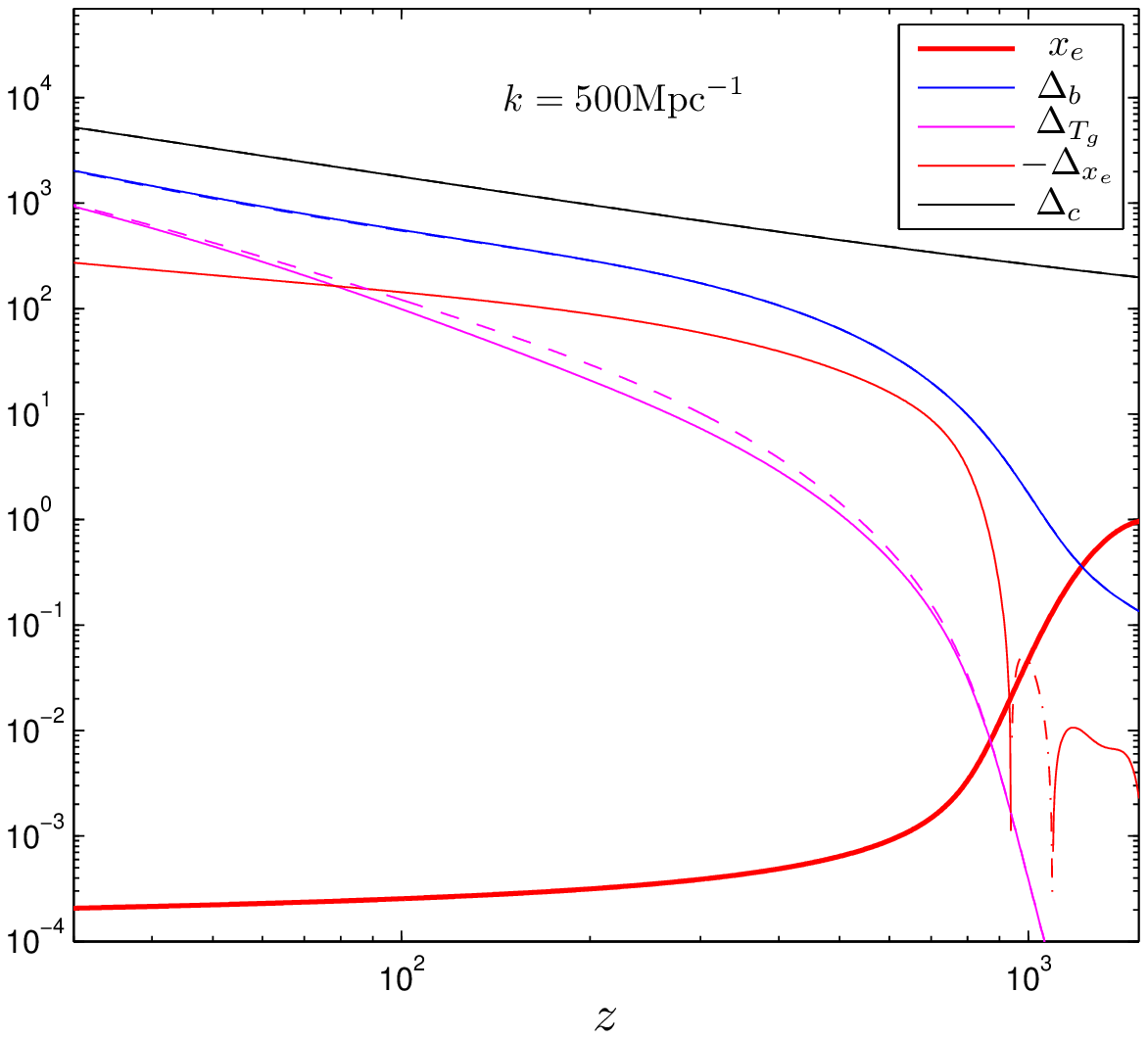,width=8cm}
\caption{Evolution of the ionization fraction, matter and gas temperature perturbations with (solid) and without (dashed) accounting for ionization fraction perturbations. Perturbations are normalized to correspond to unit initial super-horizon curvature perturbation. The ionization fraction perturbation can be positive (shown dash-dotted) during the main recombination event since overdensities can effectively delay recombination. The late-time residual local ionization fraction measured by $\Delta_{x_e}$ is insensitive to the local  timing of the main earlier recombination.
\label{evolve}
}
\end{center}
\end{figure}

After recombination the gas and CMB temperature remain coupled for a while due to Compton scattering between CMB photons and residual free electrons in the gas. However after some expansion the density and ionization fractions become lower and the gas starts to decouple from the CMB and cool adiabatically.
In detail, assuming purely Compton cooling, the background gas temperature evolves according to~\cite{Weymann65,Seager:1999km}
\begin{equation}
\dot{\bar{T}}_g + 2 \clh \bar{T}_g  = -\frac{8 a \sigma_T \bar{\rho}_\gamma x_e}{3m_e c (1+n_{\text{He}}/n_H+x_e)}\left(\bar{T}_g - \bar{T}_\gamma\right),
\end{equation}
where $\sigma_T$ is the Thomson scattering cross-section, $n_{\text{He}}$ is the number density of Helium, $T_\gamma$ is the CMB temperature, $m_e$ is the electron mass and $c$ is the speed of light. Gas temperature perturbations at wavenumber $k$ then evolve with
\begin{multline}
\dot{\Delta}_{T_g} =
 \frac{2}{3}\dot{\Delta}_c  - \frac{2}{3} k v
 -\frac{8 a \sigma_T \bar{\rho}_\gamma x_e}{3 m_e c(1+n_{\text{He}}/n_H+x_e)}\\
 \times \biggl[ \left(1-\frac{\bar{T}_\gamma}{\bar{T}_g}\right) \left\{4\Delta_{T_\gamma} + \frac{\Delta_{x_e}}{1+n_H x_e/(n_H + n_{\text{He}})}\right\}
\\+ \frac{\bar{T}_\gamma}{\bar{T}_g} \left(\Delta_{T_g} - \Delta_{T_\gamma}\right)
\biggr] ,
\end{multline}
where we neglected helium fraction perturbations, $v$ is the baryon velocity, and $\Delta_c$ is the fractional cold dark matter perturbation. This equation depends explicitly on the ionization fraction perturbation $\Delta_{x_e}$, which is often neglected. For example Ref.~\cite{Barkana:2005xu} neglects the perturbations entirely, while Ref.~\cite{Naoz:2005pd} gives an argument for them being much smaller than the density perturbations. This is true, but the effect on the temperature evolution can still be significant if not large. Ionization fraction perturbations are modelled approximately in Ref.~\cite{Furlanetto:2006wp} for the case of decaying and annihilating dark matter where the effects become large, but the impact on the standard CDM-scenario was not quantified.


An overdensity has positive $\Delta_{T_g}$ but recombines more fully than the background and hence has negative $\Delta_{x_e}$; the additional effect of the ionization fraction perturbation is therefore to reduce the coupling to the CMB and hence slightly decrease the positive temperature perturbation in an overdensity. In addition the linear baryon density perturbations are slightly modified

\section{Baryon evolution}
\label{baryons}

\begin{figure}[t!]
\begin{center}
\epsfig{figure=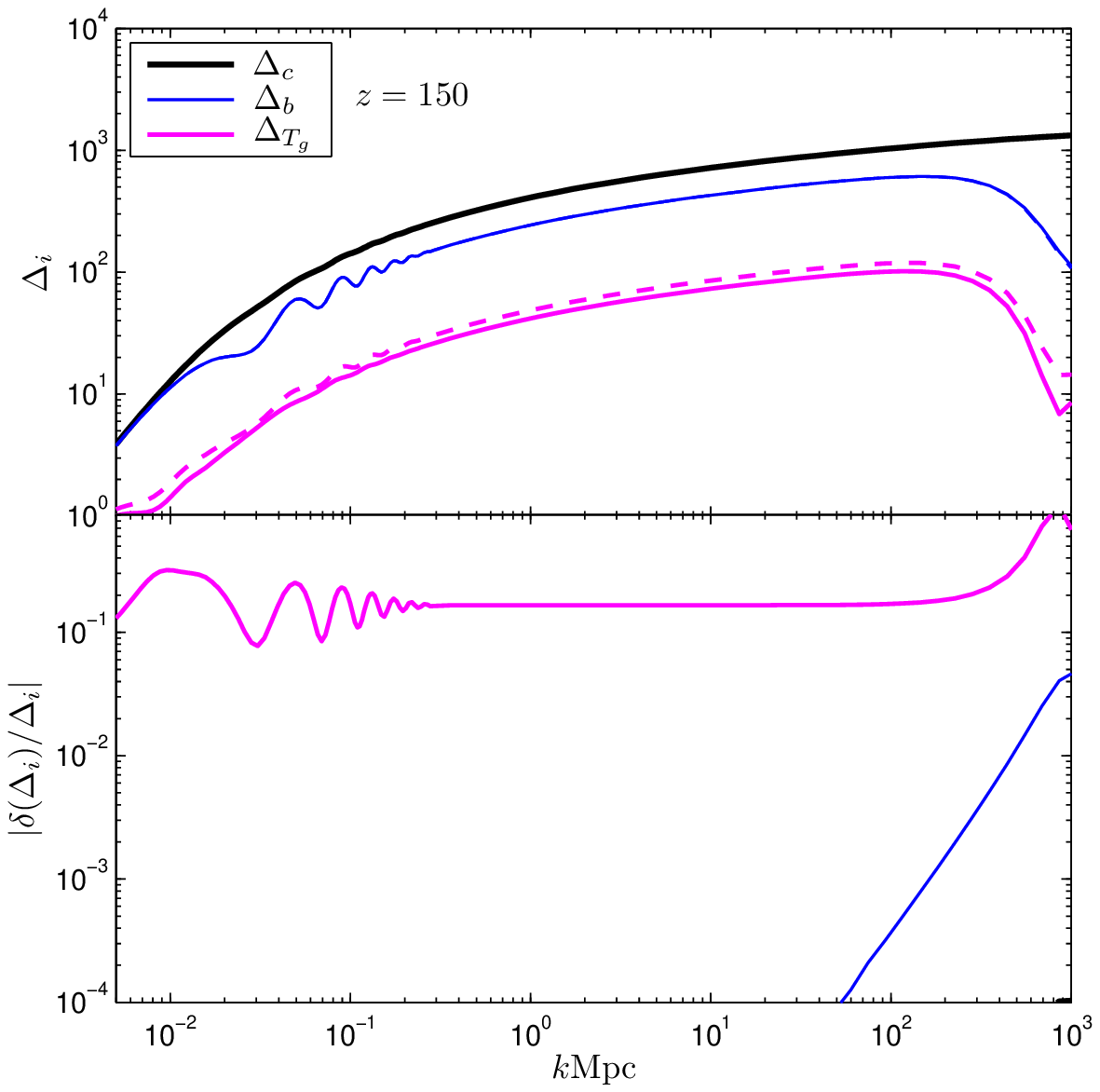,width=8cm}
\epsfig{figure=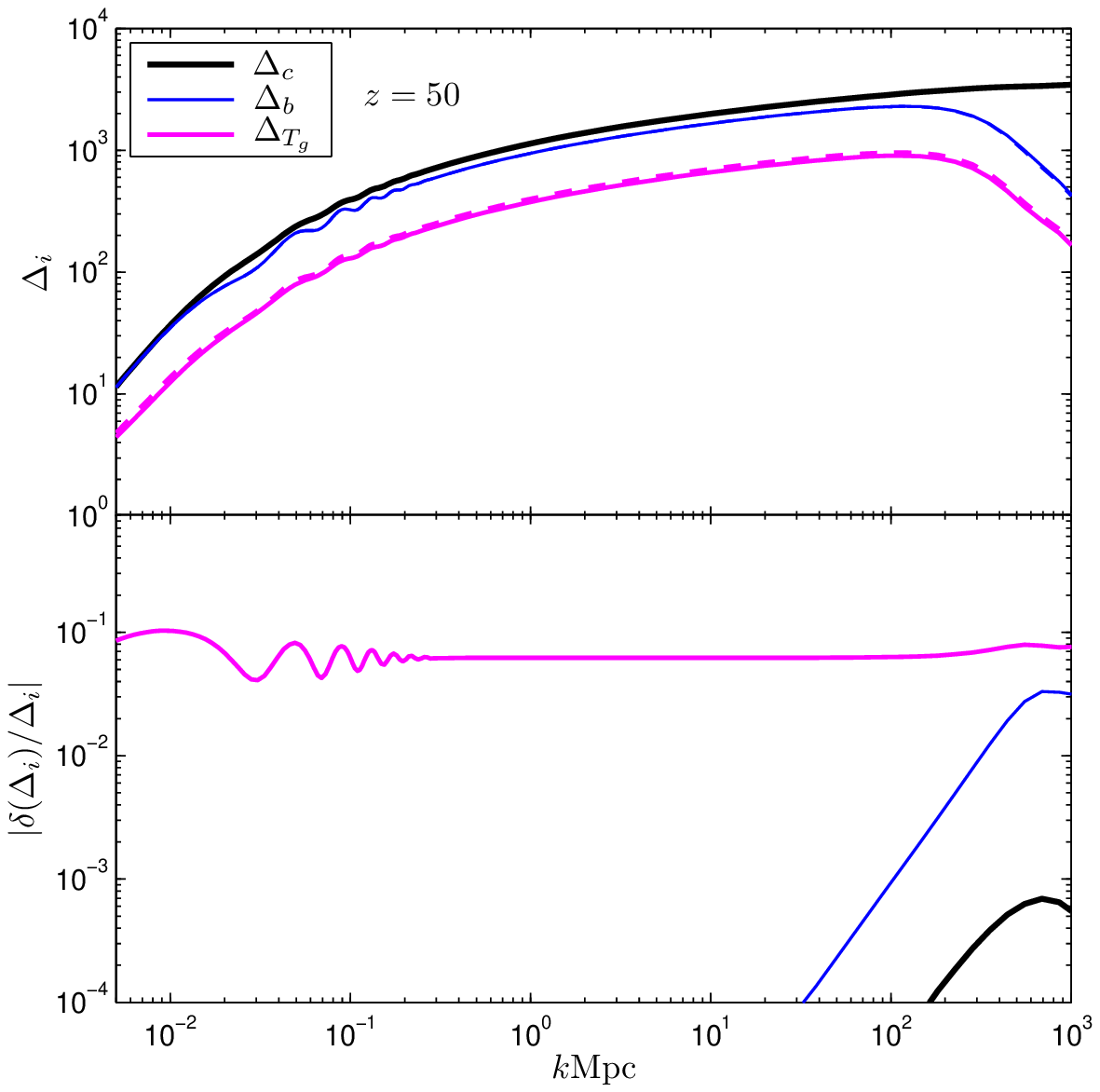,width=8cm}
\caption{Linear-theory transfer functions at redshifts $z=150$ (top plot) and $z=50$ (bottom plot) for the density and temperature perturbations (top panel), and the fractional difference due to inclusion of linear ionization fraction perturbations (bottom panel). Dashed lines show the result without including ionization fraction perturbations. Perturbations are normalized to correspond to unit initial super-horizon curvature perturbation. Note that non-linear effects will also be important on small scales at redshift $z=50$.\label{transfer}}
\end{center}
\end{figure}

After recombination Compton scattering has no significant impact on the baryon density evolution, and the gas pressure is low so that terms involving $p_g/\rho_g \ll 1$ can be neglected. The fractional synchronous gauge baryon density perturbation $\Delta_g$ then obeys the equation
\begin{eqnarray}
\ddot{\Delta}_g + \clh\dot{\Delta}_g + k^2 c_s^2 \Delta_g &=&  4\pi Ga^2 \sum_i (\delta\rho_i + 3\delta p_i)
,
\label{baryon_evolve}
\end{eqnarray}
where the sum is over the perturbations in the density and pressure of the fluid components, just baryons and dark matter during matter domination. The baryon sound speed $c_s$ is defined in terms of baryon density and pressure perturbations as $c_s^2 \equiv \delta p_b /\delta \rho_b$.
Assuming an ideal gas so that the gas pressure is given in terms of the density by $p_g \propto T_g \rho_g$, the gas pressure perturbation $c_s^2 \Delta_g=\delta p_g/\bar{\rho}_g$ is given by~\cite{Naoz:2005pd}
\begin{eqnarray}
c_s^2 \Delta_g &=& \frac{k_B \bar{T}_g}{\mu}\left(\Delta_g + \Delta_{T_g}\right) \nonumber\\
&\approx& \frac{k_B \bar{T}_g}{m_p}\left(\Delta_g + \Delta_{T_g}\right)\nonumber\\
&&\times  \left([1+x_e](1-Y_\He) + Y_\He m_p/m_\He\right),
\label{sound_speed}
\end{eqnarray}
where $\mu$ is the mean particle mass and $Y_\He$ is the mass fraction in helium. This result must be used on scales where the baryon pressure is important~\cite{Naoz:2005pd}; the common approximation that $c_s^2 \approx \ud p /\ud \rho$ is not accurate and also misses the linear effect of ionization fraction perturbations. The small-scale baryon perturbation growth is suppressed by baryon pressure, so the temperature perturbation, which depends on the ionization fraction perturbation, must be calculated properly for accurate results on scales where $k c_s/\clh \agt 1$ just after recombination.

Note that there may be important additional non-linear effects on the baryon evolution on ultra-small scales at recombination, e.g. see Refs.~\cite{Liu:2000dy,Singh:2001ub} (though note that these papers incorrectly define  $c_s^2 \equiv \dot{p} /\dot{\rho}$ in the linear baryon evolution equation following Ref.~\cite{Ma:1995ey}). In addition late-time non-linear evolution can be non-negligible on very small scales even at high redshift.

\section{Numerical results}

Numerical results for the gas temperature and baryon density can easily be obtained using perturbation evolution equations implemented in standard Boltzmann codes such as CMBFAST~\cite{Seljak96} and CAMB~\cite{Lewis:1999bs}. As of 2006 neither code correctly uses Eq.~\eqref{sound_speed} for the sound speed, largely because the error is harmless on currently observed scales.  We have modified the CAMB code to do this correctly, including approximate ionization and gas temperature perturbation evolution equations as discussed above\footnote{Code publicly available at \url{http://camb.info/sources/}}. For the ionization fraction perturbations we simply use a full set of perturbed RECFAST equations (given in the Appendix of Ref.~\cite{Lewis:2007kz}); these may not be accurate during recombination due to velocity escape-probability (and other) effects, but we have checked that the later-time dark-age ionization fraction perturbation is insensitive to these details. For redshifts $z\alt 700$ the result agrees well on most scales with integrating Eq.~\eqref{dxe} from the start of recombination. Accuracy is likely to be limited by the accuracy of the background recombination model.

Figure~\ref{evolve} shows the evolution of the various perturbations with time. For modes with wavenumber $k=1 \Mpc^{-1}$ the baryon pressure has almost no effect on the density perturbation evolution, but at $k=500\Mpc^{-1}$  pressure suppresses the baryon perturbation growth. On both scales ionization fraction perturbations have a many-percent effect on the evolution of the gas temperature fluctuations at redshifts $z \agt 40$ where the Compton coupling is not negligible. At $40\alt z\alt 100$ the fractional change in $\Delta_{T_g}$ is 5--10\%.

Collisional coupling of the hydrogen 21cm spin temperature to the gas temperature means that changes in the gas temperature affect the dark-age 21cm fluctuations on all scales (for details see Ref.~\cite{Lewis:2007kz}). At redshifts $40 \alt z \alt 70$ overdensities have a higher collisional coupling rate than the background and hence have a spin temperature closer to the gas temperature. Allowing for ionization fraction perturbations decreases the gas temperature in an overdensity and hence lowers the spin temperature further, giving a larger absorption signal. At $z \alt 40$ and the collisional spin-coupling rate falls rapidly with temperature; this can be seen as a change in the slope of the collisional de-excitation rate as a function of $T_g$ for $T_g \alt 30\K$ (see e.g. Ref.~\cite{Furlanetto:2006jb}). Here lowering the gas temperature due to ionization fraction perturbations leads to a lower collisional coupling rate, which actually causes the spin temperature to slightly increase, giving a smaller absorption signal than when $\Delta_{x_e}$ is neglected. The effect of ionization fraction perturbations perturbations is therefore to increase the signal at $z \agt 40$ and decrease it at $z \alt 40$. The 21cm power spectrum is modified at the $\agt 2\%$-level at $z\agt 50$, but the effect at $z\sim 40$ is very small due to the cancellation of the competing processes.

Figure~\ref{transfer} shows the perturbation transfer functions as a function of wavenumber. Ionization fraction perturbations affect the gas temperature transfer function on all scales at the many-percent level.
The effect on the baryons is small, but at the percent-level on scales where baryon pressure is important (i.e. modes inside the baryon sound horizon at recombination; for further discussion of baryon perturbation evolution see e.g. the Appendix of Ref.~\cite{Lewis:2007kz}). Ionization fraction perturbations reduce the temperature in overdensities which reduces the pressure, giving a small increase in the baryon density due to the smaller pressure support. The baryon density perturbations gravitationally influence the dark matter perturbations, though only at the $\clo(10^{-3})$-level, which is well below corrections due to non-linear evolution on these scales. At later times the gas cools further and the baryon perturbations relax to follow the dark matter; the late-time linear-theory baryon perturbation becomes close to what it would have been without ionization fraction perturbations.

\section{Conclusions}

Perturbation in the dark-age ionization fraction can have a significant effect on the gas temperature evolution. This must be modelled for accurate calculation of the dark-age 21cm power spectrum. The effect on the 21cm power spectrum is at the $2\%$-level on all scales at redshift $z\sim 50$ (with a larger effect on very small scales), as discussed in more detail in Ref.~\cite{Lewis:2007kz}. Baryon evolution is also affected at the percent level on scales smaller than the baryon sound horizon at recombination, corresponding to wavenumbers larger than a few hundred inverse megaparsecs. Multi-fluid numerical simulations may need to use an accurate model of fluctuations in the gas temperature to achieve accurate results on galactic scales from a given set of cosmological parameters.
\\
\section{Acknowledgements}
AL acknowledges a PPARC Advanced Fellowship and thanks Matias Zaldarriaga and Anthony Challinor for discussion and Lindsay King for comments.

\bibliography{../antony,../cosmomc}
\end{document}